\documentclass[journal,a4paper,onecolumn,12pt,draftcls,twoside]{IEEEtranTCOM}
\normalsize

\usepackage{cite}
\usepackage[cmex10]{amsmath}
\usepackage{array}
\usepackage{url}
\usepackage{amsthm}
\usepackage{amssymb}
\usepackage[T1]{fontenc}
\usepackage[utf8]{inputenc}
\usepackage[english]{babel}
\usepackage{graphicx}
\usepackage{xcolor}

\usepackage{tikz}
\usetikzlibrary{positioning,chains,fit,calc,shapes,arrows,decorations.pathreplacing}

\newtheorem{definition}{Definition}

\newtheorem{theorem}{Theorem}
\newtheorem{lemma}{Lemma}
\newtheorem{corollary}{Corollary}

\newcommand{\F}{\mathbb{F}}
\newcommand{\Fp}{\mathbb{F}_p}

\newcommand{\prob}{\mathcal{P}}
\newcommand{\R}{\mathbb{R}}
\newcommand{\Rn}{\mathbb{R}^n}

\newcommand{\vor}{\mathcal{V}}

\newcommand{\zero}{\mathbf{0}}
\newcommand{\Z}{\mathbb{Z}}
\newcommand{\Zn}{\mathbb{Z}^n}

\DeclareMathOperator*{\diff}{d}

\DeclareMathOperator*{\enc}{enc}

\DeclareMathOperator*{\snr}{SNR}

\DeclareMathOperator*{\vnr}{VNR}
\DeclareMathOperator*{\vol}{Vol}

\begin{document}
%
% paper title
% can use linebreaks \\ within to get better formatting as desired
\title{Leech Constellations\\of Construction-A Lattices}
%
%
% author names and IEEE memberships
% note positions of commas and nonbreaking spaces ( ~ ) LaTeX will not break
% a structure at a ~ so this keeps an author's name from being broken across
% two lines.
% use \thanks{} to gain access to the first footnote area
% a separate \thanks must be used for each paragraph as LaTeX2e's \thanks
% was not built to handle multiple paragraphs
%

\author{Nicola~di~Pietro and~Joseph~J.~Boutros,~\IEEEmembership{Senior~Member,~IEEE}% <-this % stops a space
  \thanks{Manuscript submitted to IEEE Transactions on Communications on December 23, 2016; revised May 21, 2017; accepted July 25, 2017. The research work presented in this paper was supported by QNRF, a member of Qatar Foundation, under NPRP project 6-784-2-329.}%
  \thanks{Nicola di Pietro is with CEA, LETI, MINATEC, F-38054, Grenoble, France (e-mail: nicola.dipietro@cea.fr).}%
  \thanks{Joseph  J.\ Boutros  is with  the  Department  of  Electrical  and  Computer  Engineering,  Texas  A\&M
    University  at  Qatar,  c/o  Qatar  Foundation,  Education  City,  Doha,  Qatar,  P.O.\ Box  23874  (e-mail: boutros@ieee.org).}%
  %\thanks{\textcolor{blue}{Results partially presented at ITA 2017, without appearing in the proceedings}}
}% <-this % stops a space

% note the % following the last \IEEEmembership and also \thanks - 
% these prevent an unwanted space from occurring between the last author name
% and the end of the author line. i.e., if you had this:
% 
% \author{....lastname \thanks{...} \thanks{...} }
%                     ^------------^------------^----Do not want these spaces!
%
% a space would be appended to the last name and could cause every name on that
% line to be shifted left slightly. This is one of those "LaTeX things". For
% instance, "\textbf{A} \textbf{B}" will typeset as "A B" not "AB". To get
% "AB" then you have to do: "\textbf{A}\textbf{B}"
% \thanks is no different in this regard, so shield the last } of each \thanks
% that ends a line with a % and do not let a space in before the next \thanks.
% Spaces after \IEEEmembership other than the last one are OK (and needed) as
% you are supposed to have spaces between the names. For what it is worth,
% this is a minor point as most people would not even notice if the said evil
% space somehow managed to creep in.

% The paper headers
\markboth{}%
{}
% The only time the second header will appear is for the odd numbered pages
% after the title page when using the twoside option.
% 
% *** Note that you probably will NOT want to include the author's ***
% *** name in the headers of peer review papers.                   ***
% You can use \ifCLASSOPTIONpeerreview for conditional compilation here if
% you desire.

% If you want to put a publisher's ID mark on the page you can do it like
% this:
%\IEEEpubid{0000--0000/00\$00.00~\copyright~2007 IEEE}
% Remember, if you use this you must call \IEEEpubidadjcol in the second
% column for its text to clear the IEEEpubid mark.

% make the title area
\maketitle
\pagestyle{plain}
\thispagestyle{plain}

%\vspace{-1cm}
\begin{abstract}
The problem  of communicating over  the additive white  Gaussian noise
(AWGN)   channel   with   lattice   codes   is   addressed   in   this
paper. Theoretically, Voronoi constellations have proved to yield very
powerful  lattice codes  when the  fine/coding lattice  is AWGN-good  and the
coarse/shaping lattice has an optimal shaping gain. However, achieving Shannon
capacity with  these premises  and practically  implementable encoding
algorithms is in general not an easy task. In this work, a new way to encode and demap Construction-A Voronoi lattice codes is presented. %The fine lattice is built via nonbinary Construction A and the coarse lattice is a direct sum of copies of the same small-dimensional lattice. This encoding and demapping method works for every Construction-A coding lattice and is independent from the choice of the small-dimensional lattice used to build the shaping lattice.
  As a meaningful application of this scheme, the second part of the paper is focused on \emph{Leech constellations} of low-density Construction-A (LDA) lattices: LDA Voronoi lattice codes are presented whose numerically measured waterfall region is situated at less than 0.8~dB from  Shannon capacity. These LDA lattice codes are based on dual-diagonal nonbinary low-density parity-check codes. With this choice, encoding, iterative decoding, and demapping have all linear complexity in the blocklength.
\end{abstract}
% IEEEtran.cls defaults to using nonbold math in the Abstract.
% This preserves the distinction between vectors and scalars. However,
% if the journal you are submitting to favors bold math in the abstract,
% then you can use LaTeX's standard command \boldmath at the very start
% of the abstract to achieve this. Many IEEE journals frown on math
% in the abstract anyway.

% Note that keywords are not normally used for peerreview papers.

% For peer review papers, you can put extra information on the cover
% page as needed:
% \ifCLASSOPTIONpeerreview
% \begin{center} \bfseries EDICS Category: 3-BBND \end{center}
% \fi
%
% For peerreview papers, this IEEEtran command inserts a page break and
% creates the second title. It will be ignored for other modes.
\IEEEpeerreviewmaketitle

\section{Introduction}
% The very first letter is a 2 line initial drop letter followed
% by the rest of the first word in caps.
% 
% form to use if the first word consists of a single letter:
% \IEEEPARstart{A}{demo} file is ....
% 
% form to use if you need the single drop letter followed by
% normal text (unknown if ever used by IEEE):
% \IEEEPARstart{A}{}demo file is ....
% 
% Some journals put the first two words in caps:
% \IEEEPARstart{T}{his demo} file is ....
% 
% Here we have the typical use of a "T" for an initial drop letter
% and "HIS" in caps to complete the first word.

% You must have at least 2 lines in the paragraph with the drop letter
% (should never be an issue)

\IEEEPARstart{D}{uring}  the   last  forty  years,  the   problem  of
transmitting digital  information via lattice constellations  has been
extensively studied, mainly for the  interest that lattices arise when
dealing with continuous channels~\cite{Forney1984,Zamir2014}; interest  that is lasting with time,
since lattice codes  may  play a role  in
physical-layer   network  coding   in  communication   networks  under
standardization~\cite{Ma2015}.

We can divide  the results on Euclidean lattice codes into  two main groups:
%as observed  in the literature:
the information-theoretical  ones, aimed
to  analytically prove  the capacity-achieving  properties of  lattice
codes~\cite{deBuda1989,Loeliger1997,Poltyrev1994,Urbanke1998,Erez2004,Nazer2011,Ingber2013,Ling2014,Ordentlich2016};
and  coding   results,  in   which  authors  design   lattice  families
particularly adapted to fast  and efficient decoding with satisfactory
performance~\cite{Sadeghi2006,Sommer2008,Sakzad2011,diPietro2012,Sadeghi2013,Yan2014,Boutros2015}. At
the price of some technical challenges  - whose solution is not always
straightforward - this second group  focuses on translating
to  the Euclidean  space the  techniques used  for designing  effective
iteratively decodable error-correcting codes  over finite fields, like
turbo  codes, low-density parity-check (LDPC) codes,  and  polar codes.  Information-theoretical
analyses exist for some lattice  families in the second group, aiming
at  establishing  a  unified  theory  that  surpasses  this  dichotomy~\cite{Vatedka2017,Yan2014,diPietro2016}.

Most  of  the available  practical  results  treat the  properties  of
lattices as  infinite constellations,  comparing performance  with the
theoretical limits established in~\cite{Poltyrev1994,Ingber2013}. When
we move our attention to  finite constellations, the theory tells that
a  winning   strategy  to  achieve   capacity  is  to   carve  Voronoi
constellations~\cite{Conway1983}  out   of  Poltyrev-limit-achieving
infinite  constellations, using  shaping  lattices that  are good  for
quantization~\cite{Erez2004,Nazer2011,diPietro2016}.   However,   in
practice  this  approach  is  hard  to  realize,  mainly  due  to  the
complexity   of   quantization   algorithms  for   general   lattices.
Erez and ten Brink proposed a scheme which employs trellis shaping to design constellations with good shaping gain and close-to-capacity performance~\cite{Erez2005bis}. Other implementable shaping  schemes have  been proposed~\cite{Ferdinand2016} and mainly applied to low-density lattice codes (LDLC)~\cite{Kurkoski2009,Sommer2009,Ferdinand2014}. Very recently, the
problem  of  encoding  constellations   of  nested  lattices  with  an
application-oriented    approach     was    treated     by    Kurkoski~\cite{Kurkoski2016}.

In this work, we focus on  Voronoi constellations for the additive white Gaussian noise (AWGN) channel
where the  fine lattices  are nonbinary Construction-A  lattices.  We
have chosen Construction A because  it constitutes a powerful tool to
build both  Poltyrev-limit-achieving  lattices  and optimal  or
near-optimal                      shaping                      regions~\cite{Loeliger1997,Erez2004,Erez2005,diPietro2016}.       Furthermore,
Construction~A  yields integer  lattices, whose encoding  and decoding
algorithms  are more  easily implemented  with respect  to noninteger
lattice constellations.

The construction of a Voronoi lattice code is divided into two main steps: first, finding a complete set of representatives of the quotient group defined by the coding lattice modulo the shaping lattice; second, reducing the representatives modulo the shaping lattice (using lattice quantization) to find the ones with the smallest norm. These points lie by construction in the Voronoi region of the shaping lattice and form the Voronoi lattice code.

The main theoretical novelty of this paper is Lemma \ref{lem:classes}, which deals with the first of the two steps above. It characterizes a specific set of representatives of the quotient group, when the coding lattice is Construction-A and the shaping lattice is contained in $p\Zn$. We apply Lemma~\ref{lem:classes} to describe a new scheme to encode and demap Construction-A lattice codes.

In the second part of the paper, we show how our method works when the shaping lattice consists of the direct sum of scaled copies of the Leech lattice. For this reason, we call the resulting lattice codes \emph{Leech constellations}. We learn from~\cite[Chap.~12]{Conway1999} or~\cite[Theorem~4.1]{Ebeling2013} that the Leech lattice is the unique (up to isomorphism) even unimodular lattice in $\R^{24}$ without vectors of Euclidean square norm $2$. The Leech lattice has numerous interesting properties and has been extensively studied~\cite{Conway1999}. Its very peculiar structure allows a deep and complete algebraic investigation. %Conway and Sloane's ``bible'' on lattices~\cite{Conway1999} is a very precious and comprehensive reference for a reader who wants to discover the Leech lattice in more detail. 
%Among all results, we would like to mention the very recent work by Cohn \emph{et al.}\~\cite{Cohn2017}, who state that the Leech lattice
In dimension $24$, it is the best known quantizer~\cite[p.~61]{Conway1999} and it corresponds to the densest sphere-packing, among both lattice and nonlattice constructions~\cite{Cohn2017}.
%\textcolor{red}{REMOVE:} The beauty of the mathematical theory on the Leech lattice is even more shining if we consider that in dimensions different from $24$ (except for the very small ones), analogous constructions are not known and similar results are even hardly conjecturable

Using direct sums of small-dimensional lattices to build the shaping lattice for Voronoi constellations is a well-known technique~\cite{Ferdinand2016,Kurkoski2016}. In this way, the high-dimensional shaping lattice inherits the same shaping gain of the small-dimensional lattices, while the complexity of the quantization operation needed for shaping remains algorithmically manageable. The scope of this paper is not to focus on the choice of the shaping lattice, but rather on the encoding and demapping scheme. Therefore, in Leech constellations we fix the small-dimensional lattices used for shaping to be (properly scaled) copies of the Leech lattice, whose shaping gain is $1.03$~dB, only $0.50$~dB  away from  the  optimal  shaping  gain of  a  spherical infinite-dimensional shaping region. The Leech lattice can be substituted with any other integer lattice without changing the theoretical description of our scheme.

To build efficiently encodable and decodable Leech constellations in very high dimensions, we cut them out of \emph{dual-diagonal} low-density  Construction-A (LDA)
lattices. To the best of our knowledge, dual-diagonal LDPC codes were
% this specific construction was
never  employed in  the  nonbinary  case and as  base  element  for
Construction  A. We  have chosen  LDA lattices  for our
Leech constellations because of the decoding performance that
they showed in our previous  work, from both the  theoretical and
practical  point   of  view~\cite{diPietro2012,diPietro2016}.   This
performance  is confirmed  by  the infinite-constellation  simulations
shown  in  Fig.~\ref{fig:infinite_dd_lda}  on the  unconstrained  AWGN
channel. Moreover  and  very  importantly,  the  family  of  dual-diagonal  LDA
lattices is designed on purpose to allow fast encoding,
%In particular, the encoding of the nonbinary LDPC code
%underlying  the LDA  construction
which can  be performed via the  chain in the
parity-check matrix.
%\textcolor{blue}{This yields a complexity linear in the lattice dimension.}
%\textcolor{blue}{The
%encoding and demapping  algorithms are described in  detail in Section
%\ref{sec:encoding_demapping}.
%;    they     are    based     on    Lemma
%\ref{lem:classes}, which is the theoretical novelty on which this work
%is based. 
Experimentally,  we  reach low error rates at a  distance of only $0.8$~dB from
Shannon  capacity (cf. Fig.~\ref{fig:leech_constellation}). To the best of our knowledge, it is the first time that
numerical results of decoding finite nonhypercubic constellations of LDA lattices
are published.

The paper is structured as follows. Section \ref{sec:lattices_awgn} recalls some knowledge
about  lattices and  lattice codes  for the  AWGN channel. In Section \ref{sec:enc_demap_constrA}, we give the general description of our new encoding and demapping scheme for Construction-A lattices, based on Lemma \ref{lem:classes}. Section \ref{sec:leech_constellations} defines Leech constellations and Section \ref{sec:dual_diag_ldpc} introduces nonbinary
dual-diagonal LDA  lattices. The latter are defined by the means  of their
parity-check matrix and are  used in  Section \ref{sec:simulations}
for numerical  simulations. The    paper   finishes    with    Section
\ref{sec:conclusion}, which  summarizes our achievements  and contains
some conclusive remarks.

%%--------------------------------------------------------------------

\subsection{Notation}
In this paper we use the bold type for vectors in row convention: $\mathbf{x} = (x_1, \ldots, x_n)$; a general coordinate of a vector is indicated by $x_i$. Capital bold letters are used for matrices and their entries are written in most cases in lower case with double index; e.g., $\mathbf{G} = \{g_{i,j}\}$. Calligraphic capital letters indicate sets: $\mathcal{B}$. The notation $O(f(n))$ indicates a function whose absolute value is upper bounded by $af(n)$ for some positive constant $a$ and for all $n$ big enough.

\section{Lattices for the AWGN channel}
\label{sec:lattices_awgn}

The scope of this section is to recall the definitions that we will need throughout the paper and fix some notation. For more details on lattices, we refer the reader to~\cite{Conway1999,Ebeling2013,Zamir2014}. 

Lattices are $\Z$-modules in the Euclidean space $\Rn$ or, equivalently, discrete additive subgroups of $\Rn$. These two definitions correspond to the following: let $\mathcal{B} = \{\mathbf{b}_1,\mathbf{b}_2,\ldots,\mathbf{b}_n\}$ be a basis over $\R$ of $\Rn$ as a vector space, then a lattice $\Lambda$ is the set of all possible linear combinations of the $\mathbf{b}_i$'s with integer coefficients; if $\mathbf{G}$ is the $n \times n$ matrix whose rows are the $\mathbf{b}_i$'s, then 
%\begin{equation*}
$\Lambda = \{\mathbf{x} \in \Rn : \mathbf{x} = \mathbf{z}\mathbf{G}, \exists \mathbf{z} \in \Zn\}$. 
%\end{equation*}
In this setting, $\mathcal{B}$ is called a \emph{basis} of the lattice, $\mathbf{G}$ a \emph{generator matrix}, and the quantity $\vol(\Lambda)=|\det(\mathbf{G})|$ is called the \emph{volume} of the lattice. It can be shown that $\vol(\Lambda) = \vol(\vor(\Lambda))$, where $\vor(\Lambda)$ is the \emph{Voronoi region} of the lattice:
\begin{equation*}
  \vor(\Lambda) = \{\mathbf{y} \in \Rn : \Vert \mathbf{y} \Vert \leq \Vert \mathbf{y} - \mathbf{x} \Vert, \forall \mathbf{x} \in \Lambda \smallsetminus \{\zero\}\}.
\end{equation*}
It is very well known that although a single lattice has (infinitely) many different bases, its volume is a characterizing invariant. Notice that we have restricted our definition to \emph{full-rank} lattices, i.e., those which have $n$ independent generator vectors in an $n$-dimensional space. We do not need to treat lower-rank lattices for the purposes of this work.

A famous and useful way to construct lattices is \emph{Construction A}~\cite{Conway1999}; this method consists in embedding into $\Rn$ an infinite number of copies of a linear code over a finite field, in a way that preserves linearity. More precisely, let $C = C[n,k]_p \subseteq \Fp^n$ be a linear code over the prime field $\F_p$ of dimension $k$ and length $n$. Identifying $\Fp^n$ with its image via an embedding $\Fp^n \hookrightarrow \Zn$, we define the lattice obtained by Construction A from $C$ as
%\begin{equation*}
  $\Lambda = \{\mathbf{x} \in \Rn : \mathbf{x} \equiv \mathbf{c} \bmod p, \exists \mathbf{c} \in C\}$.
%\end{equation*}  
Equivalently, we can write that
\begin{align*}
  \Lambda & = \{\mathbf{x} \in \Rn : \mathbf{x} = \mathbf{c} + p\mathbf{z}, \exists \mathbf{c} \in C, \exists \mathbf{z} \in \Zn\} \\
  & = C + p\Zn \subseteq \Zn.
\end{align*}
%Notice that all the points of a Construction-A lattice have integer coordinates.
It is known that~\cite[Prop.~2.5.1(d)]{Zamir2014}:
\begin{equation}
  \label{eq:vol_constrA}
  \vol(\Lambda) = p^{(n-k)}.
\end{equation}
  
In this paper, we are interested in lattices as constellations of points for the transmission of information over the AWGN channel; this channel has lattice points $\mathbf{x}$ as inputs and returns $\mathbf{y} = \mathbf{x} + \mathbf{n}$, where the $n_i$ are i.i.d.\ Gaussian random variables of mean $0$ and variance $\sigma^2$. If we do not suppose any limitation for the energy of the input $\mathbf{x}$, we say that we are considering the \emph{unconstrained} AWGN channel. A seminal theorem by Poltyrev states what follows~\cite{Poltyrev1994}:
\begin{theorem}[Poltyrev]
  \label{thm:poltyrev}
  Over the unconstrained AWGN channel and for every $\varepsilon > 0$, there exists a lattice $\Lambda \subseteq \Rn$ (in dimension $n$ big enough) that can be decoded with error probability less than $\varepsilon$ if and only if
%\begin{equation*}
  %\label{eq:vnr_poltyrev}
  $\vol(\Lambda) > (\sqrt{2 \pi e \sigma^2})^n$. 
%\end{equation*}
\end{theorem}
The condition above is often written as $\vnr > 1$, where $\vnr$ indicates the so called \emph{volume-to-noise ratio} of $\Lambda$~\cite{Forney2000}:
\begin{equation}
  \label{eq:vnr}
  \vnr = \frac{\vol(\Lambda)^{\frac2n}}{2 \pi e \sigma^2}.
\end{equation}
\begin{corollary}
  \label{cor:poltyrev}
  In the set of all lattices $\Lambda$ with fixed \emph{normalized volume} $\vol(\Lambda)^{\frac2n} = \nu$, there exists a lattice that can be decoded with vanishing error probability over the unconstrained AWGN channel only if the noise variance satisfies
  %\begin{equation*}
    $\sigma^2 < \frac{\nu}{2 \pi e} = \sigma^2_{\max}$.
  %\end{equation*}
\end{corollary} 
This corollary does not add anything new to Theorem \ref{thm:poltyrev}, but it is interesting for an operational reason: the quantity $\sigma^2$ can be interpreted as the \emph{maximum tolerable noise variance} for lattices with normalized volume $\nu$ and is often called \emph{Poltyrev limit} or \emph{Poltyrev capacity}. Whenever we work with a specific family of lattices over the unconstrained AWGN channel, an important task is to show how the decoding probability of those lattices behaves for noise variances close to $\sigma^2_{\max}$. Families of lattices that have vanishing error probability for every $\sigma^2 < \sigma^2_{\max}$ are said to be \emph{good for coding} or \emph{AWGN-good} or \emph{Poltyrev-limit-achieving}. As an example, Construction-A lattices were shown to be AWGN-good~\cite{Loeliger1997,Erez2005, Ordentlich2016}; the same holds for some ensembles of LDA lattices~\cite{diPietro2013, Vatedka2017} and generalized low-density (GLD) lattices~\cite{diPietro2015}. 

If we impose to the AWGN channel input $\mathbf{x}=(x_1,\ldots,x_n)$ the \emph{power condition}
%\begin{equation*}
  $\mathbb{E} [x_i^2] \leq P$, for some $P > 0$,
%\end{equation*}
then we call the \emph{signal-to-noise ratio} of the channel the quantity
\begin{equation}
  \label{eq:snr}
  \snr = \frac{P}{\sigma^2}.
\end{equation}
It is known that the capacity of the AWGN channel is $\frac12 \log_2(1 + \snr)$ bits per dimension~\cite[p.~365]{Proakis2008}. AWGN-good lattices are essential ingredients in lattice constructions that achieve capacity of the (constrained) AWGN channel~\cite{Erez2004,Yan2014,diPietro2016}. 

A typical efficient way of building finite - hence power-constrained - sets of lattice points for the AWGN channel, called \emph{lattice codes}, is to use pairs of nested lattices to build Voronoi constellations: we say that two lattices $\Lambda$ and $\Lambda_f$ are \emph{nested} if one is included in the other, $\Lambda \subseteq \Lambda_f$. The bigger lattice (as a set) is sometimes called the \emph{fine} lattice, hence the index $f$; its sublattice $\Lambda$ is the \emph{coarse} lattice. $\Lambda$ is a subgroup of $\Lambda_f$, therefore we can consider the quotient group:
\begin{equation*}
  \Lambda_f/\Lambda = \{\mathbf{x} + \Lambda : \mathbf{x} \in \Lambda_f\}.
\end{equation*} 
The sets $\mathbf{x} + \Lambda = \{\mathbf{x} + \mathbf{z} : \mathbf{z} \in \Lambda\}$ are called the \emph{cosets} of $\Lambda$ in $\Lambda_f$~\cite{Forney2000}. In this notation, $\mathbf{x}$ is called the \emph{leader} of the coset. The group structure is such that the coset of $\mathbf{x} + \mathbf{y}$ is equal to the coset of $\mathbf{x}$ plus the coset of $\mathbf{y}$, i.e.,  
%\begin{equation*}
  $(\mathbf{x} + \mathbf{y}) + \Lambda = (\mathbf{x} + \Lambda) + (\mathbf{y} + \Lambda)$.
%\end{equation*}
Notice that with a little abuse of notation which does not lead to confusion, the ``$+$'' symbols in the previous formula represent both the addition in $\Lambda_f$ and the addition in the quotient group $\Lambda_f/\Lambda$. It is known that the cardinality of $\Lambda_f/\Lambda$ is $M = \vol(\Lambda)/\vol(\Lambda_f)$, i.e., there exist exactly $M$ different cosets.

The lattice code $\mathcal{C}$ given by the coset leaders of $\Lambda$ in $\Lambda_f$ with smallest Euclidean norm is called the \emph{Voronoi constellation (or code)} of $\Lambda_f$ with \emph{shaping lattice} $\Lambda$~\cite{Conway1983,Forney1989}. Equivalently,
%\begin{equation*}
  $\mathcal{C} = \Lambda_f \cap \vor(\Lambda)$.
%\end{equation*}
In this context, the fine lattice $\Lambda_f$ is also called the \emph{coding lattice}. From now on, we will always assume that $\Lambda \subseteq \Lambda_f$, and $\Lambda$ and $\Lambda_f$ will be for us the standard notation for respectively the coarse/shaping and the fine/coding lattice. 

From an operational point of view, building a Voronoi constellation (i.e., encoding our lattice code) consists of two main steps:
\begin{enumerate}
\item Given the coding and the shaping lattice, being able to construct all the $M$ cosets. This means to characterize a set of $M$ different coset leaders.
\item \label{pag:point2} Given the cosets, find for each one of them the coset leader of minimum Euclidean norm, which is a point of the Voronoi constellation. This is done by the \emph{quantization} operation: the \emph{quantizer} associated with the Voronoi region of $\Lambda$ is the function 
  \begin{equation}
    \label{eq:quantizer}
    \begin{aligned}
      Q_{\vor(\Lambda)} \colon & \Rn        \to \Lambda \\
      & \mathbf{y} \ \ \mapsto \arg\min_{\mathbf{z}\in\Lambda} \Vert\mathbf{z}-\mathbf{y} \Vert
    \end{aligned}.
  \end{equation}
  Hence, if $\mathbf{x}$ is a point of a given coset of $\Lambda$ in $\Lambda_f$, its coset leader with minimum norm is $\mathbf{x} - Q_{\vor(\Lambda)}(\mathbf{x}) \in \Lambda_f$.   
\end{enumerate}
Much attention has to be paid to the fact that we are using a quantizer (or nearest-neighbor decoder) of $\Lambda$ for the procedure of encoding points of $\Lambda_f$ into a Voronoi constellation~\cite{Conway1983}. Therefore, it is important to have efficient quantization algorithms for the shaping lattice. We do not only mean optimal, mathematically well-defined, or ``numerically precise''; we mainly mean of ``manageable'' complexity. Many nice theoretical Voronoi constructions, including the capacity-achieving ones~\cite{Erez2004,diPietro2016}, cannot be implemented in high dimensions because of the complexity of the associated quantizer.

In order to achieve capacity over the AWGN channel with Voronoi constellations, optimization of both the shaping lattice and the coding lattice has to be performed~\cite{Forney1998}. Namely, the coding lattice needs to be Poltyrev-limit-achieving and the shaping lattice needs a ``spherical'' Voronoi region, that is, its shaping gain has to be optimal: we call \emph{shaping gain} of $\Lambda$ the shaping gain $\gamma_s(\Lambda) = \gamma_s(\vor(\Lambda))$ of its Voronoi region~\cite{Forney1989}:
\begin{equation}
  \label{eq:shaping_gain}
  \gamma_s(\Lambda) = \frac{n\vol(\Lambda)^{1+\frac2n}}{12} \left( \int_{\vor(\Lambda)}\Vert \mathbf{x}\Vert^2 \diff \mathbf{x}\right)^{-1}.
\end{equation}
It is an established result that $\gamma_s(\Lambda) \leq \frac{\pi e}{6} \approx 1.53$~dB for every lattice $\Lambda$. A family of lattices has an optimal shaping gain if it tends to $\frac{\pi e}{6}$ when $n$ tends to infinity. A family with this property is called \emph{good for quantization}~\cite{Erez2005}. The capacity results obtained with Construction A and LDA lattices in~\cite{Erez2004,diPietro2016} are based on shaping lattices which are good for quantization and coding lattices which are Poltyrev-limit-achieving.

\section{Encoding and demapping Construction-A lattice codes}
\label{sec:enc_demap_constrA}
In this section we show a way to perform step 1) above when the coding lattice $\Lambda_f$ is built with Construction A and the shaping lattice $\Lambda$ is contained in $p\Zn$. Notice that under these premises $\Lambda$ and $\Lambda_f$ are nested because $\Lambda \subseteq p\Zn \subseteq \Lambda_f$. The following lemma characterizes the quotient group $\Lambda_f/\Lambda$ by producing an explicit set of coset leaders. After the lemma, we will describe how to map and demap information to and from lattice codewords. 
\begin{lemma}
  \label{lem:classes}
  Let %$\Gamma = \alpha \Lambda_{24}^{\oplus \ell} \subseteq \Zn$ and let $T$ be its lower triangular generator matrix:
  $\Gamma \subseteq \Zn$ be any integer lattice and let us define $\Lambda = p \Gamma \subseteq p\Zn$. Let us call $\mathbf{T}$ a lower triangular generator matrix of $\Gamma$ with $t_{i,i}>0$ for every $i$ \footnote{Such a matrix always exists: e.g., it is enough to take any generator matrix and compute its Hermite normal form~\cite{Cohen1996}.}:
  \begin{equation*}
    \mathbf{T} = \left(
    \begin{matrix}
      t_{1,1} & 0      & \cdots  & 0      \\
      t_{2,1} & t_{2,2} & \ddots  & \vdots \\
      \vdots & \ddots & \ddots  & 0      \\
      t_{n,1} & \cdots & t_{n,n-1} & t_{n,n}
    \end{matrix} 
    \right) \in \Z^{n \times n}.
%    = \alpha \begin{bmatrix}
%      G_{24}  & 0      & \cdots & 0      \\
%      0      & G_{24}  & \ddots & \vdots \\
%      \vdots & \ddots & \ddots & 0      \\
%      0      & \cdots & 0      & G_{24}
%    \end{bmatrix},
  \end{equation*}
%  where the matrix $G_{24}$ appears exactly $\ell$ times. 
  Let us call $\mathcal{S}$ the set
  \begin{equation*}
    \mathcal{S} = \{0,\ldots,t_{1,1}-1\} \times \{0,\ldots,t_{2,2}-1\} \times \cdots \times \{0,\ldots,t_{n,n}-1\}.
  \end{equation*}
  Let $\Lambda_f = C + p\Zn$ be a Construction-A lattice; we consider the usual embedding of $C$ in $\Zn$ via the coordinate-wise morphism $\Fp \hookrightarrow \{0,1,\ldots,p-1\} \subseteq \Z$, hence $C \subseteq \{0,1,\ldots,p-1\}^n$.

  Then $C + p\mathcal{S} = \{\mathbf{c} + p\mathbf{s} \in \Zn : \mathbf{c} \in C,\ \mathbf{s} \in \mathcal{S}\}$ is a complete set of coset leaders of $\Lambda_f/\Lambda$.
\end{lemma}
\begin{IEEEproof}
  In principle $|C + p\mathcal{S}| \leq |C||\mathcal{S}|$, though it is easy to show that the equality is achieved: suppose that $\mathbf{c} + p\mathbf{s} = \mathbf{d} + p\mathbf{t}$ for some $\mathbf{c},\mathbf{d} \in C \subseteq \{0,1,\ldots,p-1\}^n$ and $\mathbf{s},\mathbf{t} \in \mathcal{S}$; then $\mathbf{c} \equiv \mathbf{d} \bmod p$. This means that  $\mathbf{c} = \mathbf{d}$, because all the $c_i$'s and $d_i$'s are in $\{0,1,\ldots,p-1\}$. This in turn implies that $\mathbf{s} = \mathbf{t}$. Hence every pair $(\mathbf{c},\mathbf{s}) \in C \times \mathcal{S}$ generates a different point in $C + p\mathcal{S}$ and $|C||\mathcal{S}| = |C + p\mathcal{S}|$.

  Now, by the triangularity of $\mathbf{T}$ and the definition of $\mathcal{S}$, we have that 
  \begin{equation*}
    \vol(\Lambda) = \vol(p\Gamma) = p^n \prod_{i=1}^nt_{i,i} = p^n |\mathcal{S}|.
    \end{equation*}
  If we also take into account \eqref{eq:vol_constrA} and that the cardinality of $\Lambda_f/\Lambda$ is
  \begin{equation*}
    M = \frac{\vol(\Lambda)}{\vol(\Lambda_f)} = \frac{\vol(\Lambda)}{p^{n-k}},
  \end{equation*}
  then we easily obtain that 
  \begin{equation*}
    |C + p\mathcal{S}| = |C||\mathcal{S}| = p^k \frac{\vol(\Lambda)}{p^n} = M.
  \end{equation*}
  
  We have just proved that $C + p\mathcal{S}$ and $\Lambda_f/\Lambda$ have the same cardinality. At this point, to conclude the proof of the lemma, it is sufficient to show that any two elements of $C + p\mathcal{S}$ belong to different cosets. Equivalently, we will prove the following: if $\mathbf{x}, \mathbf{y} \in C+p\mathcal{S}$ belong to the same coset, then $\mathbf{x} = \mathbf{y}$. 
  
  Now, $\mathbf{x} = \mathbf{c} + p\mathbf{s}$ and $\mathbf{y} = \mathbf{d} + p\mathbf{t}$ are in the same coset if and only if $\mathbf{x} - \mathbf{y} = \mathbf{c} - \mathbf{d} + p(\mathbf{s} - \mathbf{t})$ is in $\Lambda \subseteq p\Zn$. This holds only if $\mathbf{c} - \mathbf{d} \in p\Zn$ and consequently only if $c_i - d_i = 0$ for every $i$, because $0$ is the only element of $p\Z$ that can be obtained by subtracting two numbers of $\{0,1,\ldots,p-1\}$. Thus $\mathbf{x}$ and $\mathbf{y}$ are in the same coset only if $\mathbf{c} = \mathbf{d}$ and $\mathbf{x} - \mathbf{y} = p(\mathbf{s} - \mathbf{t}) \in \Lambda = p\Gamma$. This implies that $\mathbf{s} - \mathbf{t} \in \Gamma$, i.e., $\mathbf{s} - \mathbf{t} = \mathbf{z}\mathbf{T}$ for some $\mathbf{z} \in \Z^n$. Let $\mathbf{U} = \{u_{i,j}\}$ be the inverse of $\mathbf{T}$: it is lower triangular and $u_{i,i} = t_{i,i}^{-1}$ for every $i$.
  The relation $(\mathbf{s} - \mathbf{t})\mathbf{U} = \mathbf{z}$ implies that
  \begin{equation*}
    \sum_{j=i}^n (s_j-t_j)u_{j,i} \in \Z \text{ for every } i.
  \end{equation*}
  When $i=n$, the condition is simply $(s_n-t_n)t_{n,n}^{-1} \in \Z$, which implies that $s_n - t_n = 0$ because by definition of $\mathcal{S}$ we have $\vert s_n-t_n \vert\leq t_{n,n} - 1 < t_{n,n}$. Using the equality $s_n = t_n$ in the case $i=n-1$, we obtain that $s_{n-1} = t_{n-1}$ too. Moving recursively backwards to $i=n-2,n-3,...,1$ and using each time the new equalities, we conclude that $s_i = t_i$ for every $i$, i.e., $\mathbf{s} = \mathbf{t}$ and, as wanted, $\mathbf{x} = \mathbf{y}$.
\end{IEEEproof}

\subsection{Encoding}
\label{sec:encoding_constrA}
Based on Lemma \ref{lem:classes}, the encoding of Voronoi constellations of Construction-A lattices with shaping lattice $\Lambda \subseteq p\Zn$ can be done as follows:
\begin{enumerate}
\item Information is represented by integer vectors of the set $\mathcal{M} = \Fp^k \times \mathcal{S}$. %(where $\mathcal{S}$ is defined in Lemma \ref{lem:classes}).
\item Let $\mathbf{m} = (\mathbf{u},\mathbf{s}) \in \mathcal{M}$ be a message to encode, with $\mathbf{u} \in \Fp^k$ and $\mathbf{s} \in \mathcal{S}$. Let $\mathbf{c}$ be the codeword of $C$ associated with $\mathbf{u}$: if $\enc_C(\cdot)$: $\Fp^k \rightarrow \Fp^n$ is an encoder of $C$, then $\mathbf{c}=\enc_C(\mathbf{u})$.
  %, if $\mathbf{G}_C$ is a generator matrix of the code $C$, then $\mathbf{c}=\mathbf{u}\mathbf{G}_C$.
\item Let $\mathbf{x}' = \mathbf{c} + p\mathbf{s} \in \Lambda_f$. By Lemma \ref{lem:classes}, any two different messages $\mathbf{m}$ correspond to two different coset leaders $\mathbf{x}'$ of $\Lambda$ in $\Lambda_f$.
\item Let $Q_{\vor(\Lambda)}(\cdot)$ be the lattice quantizer \eqref{eq:quantizer} associated with the Voronoi region of the shaping lattice $\Lambda$. Then, the message $\mathbf{m}$ is encoded to the lattice codeword 
  \begin{equation*}
    \mathbf{x} = \mathbf{x}' - Q_{\vor(\Lambda)}(\mathbf{x}') \in \mathcal{C} = \Lambda_f \cap \vor(\Lambda).
  \end{equation*}
\end{enumerate}
Notice that $\mathbf{y}$ and $\mathbf{y} - Q_{\vor(\Lambda)}(\mathbf{y})$ belong to the same coset for every $\mathbf{y} \in \Lambda_f$. This guarantees that any two different messages are indeed encoded to different points of the Voronoi constellation.

The encoding procedure that we have just described differs from what is typically done for Construction-A lattice codes (e.g., in~\cite{Kurkoski2016}) and the bijection between $\Lambda_f/\Lambda$ and $C + p\mathcal{S}$ provided by Lemma \ref{lem:classes} did not appear in the literature before this paper, to the best of our knowledge. Three features deserve to be highlighted:
  \begin{itemize}
  \item Lemma \ref{lem:classes} provides a new way to label lattice codewords: information is represented by elements of $\mathcal{M} = \mathbb{F}_p^k \times \mathcal{S}$; hence, each information point has $k+n$ coordinates, whereas lattice codewords have $n$ coordinates. This is different from all the classical representations of information for lattice codes, in which information points have the same number of coordinates of lattice points.
  \item All the $k+n$ coordinates of a message $\mathbf{m}$ can be chosen independently.
  \item Encoding is not performed via the generator matrix of the coding lattice, as typically done in the literature~\cite{Ferdinand2016,Kurkoski2016}. Lemma \ref{lem:classes} tells that we can first build codewords of the linear code $C$ and then translate them by points of $p\mathcal{S}$. This may seem just a detail in the whole process, but it is not a marginal point. Using a low-complexity encoding of $C$, this approach will allow us in Section \ref{sec:dual_diag_ldpc} to build Voronoi constellations whose encoding complexity is linear in $n$, whereas, in general, encoding via the lattice generator matrix has complexity proportional to $n^2$.
  \end{itemize}

\subsection{Demapping}
\label{sec:demapping_constrA}

In the process of communication, besides encoding and decoding a constellation point, it is necessary to specify how \emph{demapping} works, i.e., how to reobtain the information vector from a given constellation point. We describe in this subsection how to derive $\mathbf{m}=(\mathbf{u},\mathbf{s})$ back from a given lattice codeword $\mathbf{x}$. Namely, we apply the following steps:
\begin{enumerate}
\item $Q_{\vor(\Lambda)}(\mathbf{y}) \in \Lambda = p\Gamma \subseteq p\Zn$ for every $\mathbf{y} \in \Rn$, hence we can obtain $\mathbf{c}$ simply by reducing modulo $p$ the point $\mathbf{x} = \mathbf{c} + p\mathbf{s} - Q_{\vor(\Lambda)}(\mathbf{x}')$. 
\item How to derive $\mathbf{u}$ from $\mathbf{c}$ strictly depends on how codewords of $C$ are encoded. This may change case by case, depending on applications. As a general example that corresponds to what we propose in Section \ref{sec:dual_diag_ldpc}, we can suppose that the codewords of $C$ are encoded via a \emph{systematic} encoder $\enc_C(\cdot)$, so that $\mathbf{c} = \enc_C(\mathbf{u}) = (\mathbf{u}|\mathbf{c}')$, for some \emph{parity symbols} $\mathbf{c}' \in \Fp^{n-k}$. %generator matrix $\mathbf{G}_C$ of $C$.
  Hence, $\mathbf{u}$ is automatically given by the \emph{information symbols} of $\mathbf{c}$. We need now to compute $\mathbf{s}$.
\item At this point, since we know both $\mathbf{x}$ and $\mathbf{c}$, we can recover 
  \begin{equation}
    \label{eq:r}
    \mathbf{r} = \frac{(\mathbf{x} - \mathbf{c})}{p} = \mathbf{s} - \frac{1}{p}Q_{\vor(\Lambda)}(\mathbf{x}') = \mathbf{s} - \mathbf{q} \in \Zn,
  \end{equation}
  for some $\mathbf{q} \in \Gamma = p^{-1}\Lambda$. In particular, if $\mathbf{T}$ is the triangular generator matrix of $\Gamma$ as in \eqref{eq:T}, %and $U$ is its inverse, 
  $\mathbf{r} = \mathbf{s} - \mathbf{z}\mathbf{T}$ for some unknown $\mathbf{z} \in \Zn$. %and $\mathbf{z} = (\mathbf{s} - \mathbf{r})U$.
\item %Recall that $u_{i,i} = t_{i,i}^{-1}$ for every $i$; hence 
  By triangularity of $\mathbf{T}$, the $i$-th coordinate of the previous equality is:
  \begin{equation}
    \label{eq:zi}
    %z_i = (s_i-r_i)t_{i,i}^{-1} + \sum_{j=i+1}^n (s_j-r_j)u_{j,i}.
    r_i = s_i - z_it_{i,i} - \sum_{j=i+1}^n z_jt_{j,i}.
  \end{equation}
  For $i=n$, the rightmost term of \eqref{eq:zi} is absent and we have 
  \begin{equation}
    \label{eq:sn}
    s_n = r_n + z_n t_{n,n}. %\in t_{n,n}\Z. 
  \end{equation}
  $r_n$ is known from \eqref{eq:r} and we aim to find $s_n$. This is easy, because the two following conditions uniquely identify it:
  \begin{itemize}
  \item $s_n \equiv r_n \bmod t_{n,n}$;
  \item $0 \leq s_n \leq t_{n,n}-1$ (by definition of $\mathcal{S}$).
  \end{itemize}
  Furthermore, after having found $s_n$, we can compute $z_n$ from \eqref{eq:sn}.
\item For $i=n-1$, \eqref{eq:zi} yields 
  \begin{equation}
    \label{eq:sn_1}
    %s_{n-1}-r_{n-1} + t_{n-1,n-1}(s_n-r_n)u_{n,n-1} = z_{n-1}t_{n-1,n-1} \in t_{n-1,n-1}\Z.
    s_{n-1} = r_{n-1} + z_{n-1}t_{n-1,n-1} + z_n t_{n,n-1}. %\in t_{n-1,n-1}\Z.
  \end{equation}
  The two unknowns here are $s_{n-1}$ and $z_{n-1}$. In a similar way as before, $s_{n-1}$ can be explicitly computed because:
  %On the left side, the only unknown is $s_{n-1}$; we can obtain it in a similar way as before, since it is uniquely by the two conditions: 
  \begin{itemize}
  \item $s_{n-1} \equiv r_{n-1} + z_n t_{n,n-1} \bmod t_{n-1,n-1}$;
  \item $0 \leq s_{n-1} \leq t_{n-1,n-1}-1$ (by definition of $\mathcal{S}$).
  \end{itemize}
  Once we have $s_{n-1}$, it is easy to obtain $z_{n-1}$ from \eqref{eq:sn_1}.
\item Going on with the same strategy, using recursively at the $i$-th step the values of $s_j$ and $z_j$ already computed for $j = i+1,i+2,\ldots,n$, we obtain $s_i$ for the remaining $i=n-2,n-3,\ldots,1$. This concludes the demapping procedure.
\end{enumerate}

\section{Leech constellations}
\label{sec:leech_constellations}
%The main purpose of this paper is to propose a
  From now on, we will put into practice the coding scheme described in Section \ref{sec:enc_demap_constrA} with the goal of designing lattice codes with encoding and demapping complexity linear in $n$. In this section, we choose a standard solution to simplify the algorithmic problem of quantization for shaping. We will focus on Voronoi constellations in which %the fine lattice is built via Construction A and the shaping lattice is a direct sum of (scaled) copies of the Leech lattice.
  the shaping lattice $\Lambda$ is the direct sum of low-dimensional lattices: $\Lambda = \Lambda_s^{\oplus \ell}$, for some $\ell$ proportional to $n$ and some lattice $\Lambda_s \subseteq \R^{n/\ell}$. This is a standard approach, which yields a coarse lattice with the same shaping gain of $\Lambda_s$. Kurkoski~\cite{Kurkoski2016} considers this construction when the fine lattice is built via Construction A and the authors of~\cite{Ferdinand2016} use it with low-density lattice codes (LDLC).

The choice of taking $\Lambda = \Lambda_s^{\oplus \ell}$ results in a low-complexity quantizer of the shaping lattice. In particular, for every $\mathbf{y} \in \Rn$ we have:
\begin{equation*}
  Q_{\vor(\Lambda)}(\mathbf{y}) = \left(Q_{\vor(\Lambda_s)}(\mathbf{y}_1)\ |\ Q_{\vor(\Lambda_s)}(\mathbf{y}_2)\ | \cdots |\ Q_{\vor(\Lambda_s)}(\mathbf{y}_\ell)  \right),
\end{equation*}
where $Q_{\vor(\Lambda_s)}(\cdot)$ is the $n/\ell$-dimensional Voronoi quantizer of $\Lambda_s$ and, for every $i=1,2,\ldots,\ell$,
\begin{equation*}
  \mathbf{y}_i = \left(y_{1 + n(i-1)/\ell}, y_{2 + n(i-1)/\ell}, \ldots, y_{ni/\ell}\right).
\end{equation*}
Hence, applying $Q_{\vor(\Lambda)}(\cdot)$ is equivalent to apply $\ell$ independent quantizers $Q_{\vor(\Lambda_s)}(\cdot)$. When $\Lambda_s$ has constant dimension in $n$, the complexity of the quantization operation is $O(\ell) = O(n)$.

The scope of this paper is not to introduce any fundamental novelty concerning the construction of the shaping lattice. For this reason, we choose to fix it once for all: from now on, $\Lambda_s$ will be (a scaled copy of) the \emph{Leech lattice}. It is known that we need shaping lattices with a high shaping gain to obtain Voronoi constellations with decoding performance close to capacity. The Leech lattice is the best-known quantizer in dimension $24$~\cite[p.~61]{Conway1999} and has a shaping gain of about $1.03$~dB~\cite{Forney1992}. This corresponds to a difference of around $0.50$~dB from the optimal shaping gain. If we work with AWGN-good fine lattices, our experimental target is to achieve numerically measured decoding error probabilities with a waterfall region situated at around $0.50$~dB from Shannon capacity. We will show in Section \ref{sec:simulations} how close we can get to this result.%, but before, we need to start with a theoretical description of our particular Voronoi constellations.

Now, let $\Lambda_f = C[n,k]_p + p\Zn$ be the $n$-dimensional Construction-A fine lattice and let us suppose from now on that $n = 24 \ell$ for some integer $\ell$. %It is known that~\cite[Prop.~2.5.1(d)]{Zamir2014}:
%\begin{equation}
%  \label{eq:vol_constrA}
%  \vol(\Lambda_f) = p^{(n-k)}.
%  \end{equation}
Let $\mathbf{G}_{24}$ be the lower triangular generator matrix of the Leech lattice proposed by Conway and Sloane in~\cite[p.~133]{Conway1999}, but with all the coordinates multiplied by $\sqrt{8}$.
In particular, we are considering an integer version of the Leech lattice: $\mathbf{G}_{24} \in \Z^{24 \times 24}$. In spite of scaling, we can still call it without confusion \emph{the} Leech lattice and we denote it $\Lambda_{24}$. Also, one can check that
\begin{equation*}
  \det(\mathbf{G}_{24}) = \vol(\Lambda_{24}) = (\sqrt{8})^{24} = 2^{36}.
\end{equation*}

Now, consider the lattice given by the following direct sum of $\ell$ copies of the Leech lattice:
\begin{equation*}
  \Lambda_{24}^{\oplus \ell} = \{\mathbf{x} = (\mathbf{x}_1|\mathbf{x}_2|\cdots|\mathbf{x}_\ell) \in \Rn : \mathbf{x}_i \in \Lambda_{24}, \forall i\} \subseteq \Zn
\end{equation*}
and let us call $\Gamma = \alpha \Lambda_{24}^{\oplus \ell}$, for $\alpha \in \mathbb{N}\smallsetminus\{0\}$.
The generator matrix of $\Gamma$ is the $n \times n$ diagonal matrix obtained by diagonally juxtaposing $\ell$ copies of $\mathbf{G}_{24}$ multiplied by $\alpha$ (and filling with zeroes all the other entries):
\begin{equation}
  \label{eq:T}
  \mathbf{T} = \left(
  \begin{matrix}
    \alpha \mathbf{G}_{24} & 0                     & \cdots & 0                      \\
    0                     & \alpha \mathbf{G}_{24} & \ddots & \vdots                 \\
    \vdots                & \ddots                & \ddots & 0                       \\
    0                     & \cdots                & 0      & \alpha \mathbf{G}_{24}
  \end{matrix} 
  \right) \in \Z^{n \times n}.
\end{equation}
If we denote $V = (\sqrt{8})^{24}$ the volume of $\Lambda_{24}$, it is easy to compute %from \eqref{eq:L24matrix} 
that $\vol(\Gamma) = \alpha^n V^\ell$. In what follows, the shaping lattice will always be $\Lambda = p\Gamma = p\alpha \Lambda_{24}^{\oplus \ell}$.

\begin{definition}
  We call \emph{Leech constellation} of a Construction-A lattice its Voronoi constellation when the shaping lattice is a direct sum of (conveniently scaled) copies of the Leech lattice: the fine lattice is $\Lambda_f = C + p\Zn$ and the shaping lattice is $\Lambda = p\alpha\Lambda_{24}^{\oplus \ell}$.
\end{definition}
Leech constellations are well defined because
  \begin{equation*}
    \Lambda = p\Gamma = p\alpha\Lambda_{24}^{\oplus \ell} \subseteq p\Z^n \subseteq C + p\Zn = \Lambda_f.
  \end{equation*}
  If we call $g_1, g_2, \ldots g_{24}$ the diagonal elements of $\mathbf{G}_{24}$, then the set $\mathcal{S}$ of Lemma \ref{lem:classes} becomes:
\begin{equation*}
  %\label{eq:S}
  \mathcal{S} = \left( \{0,1,\ldots,\alpha g_1-1\} \times \cdots \times \{0,1,\ldots,\alpha g_{24}-1\} \right)^{\times \ell}.
\end{equation*}

We can easily compute the cardinality $M$ of the Leech constellation: if $R=k/n$ is the rate of the code $C$,
\begin{align*}
  %\label{eq:cardinality}
  M & = |\Lambda_f \cap \vor(\Lambda)| = \left| \Lambda_f / \Lambda \right| = \frac{\vol(\Lambda)}{\vol(\Lambda_f)} \\
  & = \frac{p^n \alpha^n V^\ell}{p^{n-k}} = p^k \alpha^n V^\ell = \left(p^R \alpha V^{1/24}\right)^n.
\end{align*}
Consequently, the \emph{information rate} of the Leech constellation $\mathcal{C}$ is
\begin{equation*}
  R_{\mathcal{C}} = \frac{\log_2 M}{n} % = \frac{\log_2\left(\left( p^R \alpha V^{1/24} \right)^n\right)}{n} \text{ bits/dim } \\ %= \log_2\left(\alpha p^R \sqrt{8} \right)\text{ bits/dim } \\
  %= R \log_2 p + \log_2 \alpha + \frac{1}{24} \log_2V
  = R \log_2 p + \log_2 \alpha + \frac32 \text{ bits/dim },
\end{equation*}
because $V = (\sqrt{8})^{24}$. By tuning the parameters $p,R$, and $\alpha$, we can fix different information rates. As an example, the values $\alpha=1$, $p=13$, and $R=1/3$ that are used in the simulations of Section \ref{sec:simulations}, yield a rate of $R_{\mathcal{C}} \approx 2.73$ bits per dimension.

Independently from the decoder used for $\Lambda_f$, if we apply the coding scheme of Section \ref{sec:enc_demap_constrA} to Leech constellations, we can observe the following:
\begin{itemize}
\item The complexity of the encoding algorithm resides in steps 2) and 4) of Section \ref{sec:encoding_constrA}. Because of what we pointed out at the beginning of this section, step 4) (quantization) has practical complexity, linear in $n$. The linearity constant is a power of $24$, due to the complexity of the Leech quantizer, polynomial in its dimension. Numerical simulations like the ones of Section \ref{sec:simulations} tell us that the complexity of the Leech-constellation encoder is manageable and we are capable of simulating encoding and decoding up to dimension $n=10^6+8$ (the addition of $8$ to the round number $10^6$ is needed to make $n$ divisible by $24$). Notice that these simulations use the codes that we will design in Section \ref{sec:dual_diag_ldpc}, for which also step 2) of Section \ref{sec:encoding_constrA} (encoding of $C$) is $O(n)$.
\item In general, the complexity of demapping resides in computing $s_i$ from \eqref{eq:zi} (as in \eqref{eq:sn_1} for $i = n-1$). Nevertheless, in the case of Leech constellations, for every given $i$, all but at most $24$ of the $t_{j,i}$'s are equal to zero\footnote{More precisely, with our choice of $\mathbf{G}_{24}$, there are in average $5.625$ nonzero $t_{j,i}$'s per column; the minimum is $1$ and the maximum is $21$.}. Therefore, each step from 1) to 6) of Section \ref{sec:demapping_constrA} requires a constant (in $n$) number of operations for every $i$ and the complexity of demapping is $O(n)$ too. More generally, when using copies of a lattice of dimension $d$ in the direct sum that produces the shaping lattice, the complexity of demapping is $O(dn)$.
\end{itemize}

\section{Dual-diagonal LDA lattices}
\label{sec:dual_diag_ldpc}

In Section \ref{sec:lattices_awgn}, we mentioned that we need  fine lattices with good performance (ideally Poltyrev-limit-achieving) for the construction of strong Voronoi constellations; furthermore, through Section \ref{sec:encoding_constrA} and \ref{sec:leech_constellations} we established that linear-complexity encoding of Leech constellations is possible if the encoding of the underlying $p$-ary code $C$ is linear in $n$ too. In this section, we propose the algebraic construction of a lattice family which possesses both qualities: good performance over the unconstrained AWGN channel and fast encoding. As fine lattices, we choose a particular family of \emph{LDA lattices}:
\begin{definition}
  We call a \emph{low-density Construction-A (LDA)} lattice a lattice built with Construction A when the underlying code $C$ is a low-density parity-check (LDPC) code.
\end{definition}
LDPC codes were invented by Gallager~\cite{Gallager1963}, have had a huge success, and do not need further introduction. LDA lattices were proposed by the authors of this work a few years ago~\cite{diPietro2012}; they are endowed with an iterative low-complexity decoder which allows fast decoding with satisfactory performance. Well-defined ensembles of LDA lattices were proved to be Poltyrev-limit-achieving first~\cite{diPietro2013}, then also Shannon-capacity-achieving~\cite{diPietro2016} and good for other communication-related problems~\cite{Vatedka2017}.

\begin{definition}
  A square matrix $\mathbf{A} = \{a_{i,j}\}$ is said \emph{dual-diagonal} if all its entries are equal to $0$ except for the $a_{i,i}$'s and the $a_{i,i-1}$'s.
\end{definition}

We look for LDA lattices that can be rapidly encoded. Our solution is to use Construction A with LDPC codes whose parity-check matrix $\mathbf{H}$ has a dual-diagonal submatrix. By extension, we call them \emph{dual-diagonal LDPC codes} and their associated LDA lattices \emph{dual-diagonal LDA lattices}. Recall that a parity-check matrix of a code $C$ is a matrix $\mathbf{H}$ which defines the code as: $C = \{\mathbf{x} \in \Fp^n : \mathbf{H}\mathbf{x}^T \equiv \zero \bmod p\}$. For our construction, we impose that $\mathbf{H}$ has the following structure:
\begin{equation}
  \label{eq:H}
  %\text{\Large$\mathbf{H}$} = \left(
  \mathbf{H} = \left(\mathbf{L}|\mathbf{R}\right);
\end{equation}
$\mathbf{H}$ has $n-k$ rows, $n$ columns, and its right submatrix $\mathbf{R}$ is the following square \emph{dual-diagonal} matrix:
\begin{equation*}
  %\begin{matrix}
  %  \hfill \text{\ \ \ \LARGE$\mathbf{R}$\ \ } \hfill
  %\end{matrix}
  %\, \middle \vert \,
  %\mathbf{R} =
  \left(
  \begin{matrix}
    h_{1,k+1} & 0       & \cdots & \cdots  & 0 \\  
    h_{2,k+1} & h_{2,k+2} & \ddots &            & \vdots \\
    0       & \ddots   & \ddots & \ddots     & \vdots \\
    \vdots  & \ddots   & h_{n-k-1,n-2} & h_{n-k-1,n-1} & 0 \\
    0       & \cdots   & 0 & h_{n-k,n-1}   & h_{n-k,n}
  \end{matrix}
  \right),
\end{equation*}
with $h_{i,k+i} \neq 0$ for every $i=1,2,\ldots,n-k$ and $h_{i,k+i-1} \neq 0$ for every $i=2,3,\ldots,n-k$. %The matrix $\mathbf{H}$ has $n-k$ rows and $n$ columns and its right submatrix is square and \emph{dual-diagonal}.
Moreover, to build LDA lattices, we need $\mathbf{H}$ to be sparse, hence its left submatrix $\mathbf{L}$ (of size $(n-k)\times k$) has to be sparse too. In particular, we choose it also to be \emph{regular}: it has a fixed constant number of nonzero entries in every column and row. We call these numbers respectively the \emph{column degree} $d_c$ and the \emph{row degree} $d_r$ of $\mathbf{L}$. Notice that, a priori, $\mathbf{L}$ could be taken irregular and its degree distribution could be optimized, but the standard techniques used for binary LDPC codes cannot be applied in this nonbinary context. Fine-tuning the degree distribution of $\mathbf{L}$ goes beyond the scope of this paper.

By construction, $\mathbf{H}$ is full-rank, hence the rate of the LDPC code that it identifies is $R = k/n$. Furthermore, all the rows of $\mathbf{H}$ have degree $d_r + 2$, except for the first row, that has degree $d_r+1$. If we count the number of nonzero entries at first column by column and then row by row, we relate the degrees and the code parameters via the following equality:
\begin{equation*}
  d_ck + 2(n-k-1) + 1 = (d_r+2)(n-k-1) + d_r + 1.
\end{equation*} 
By simplifying the previous formula, we can easily derive that
\begin{equation}
  \label{eq:R}
  R = \frac{k}{n} = \frac{d_r}{d_r+d_c}.
\end{equation}
This kind of dual-diagonal parity-check matrix has been used for several practical applications of binary LPDC and repeat-accumulate codes~\cite[Sec.~6.5]{Ryan2009}, %\cite{He2006,Guo2010,Schmalen2012,He2014,Li2014},
  but to the best of our knowledge it was never applied to nonbinary constructions or lattice constructions. The main advantage of using nonbinary LDPC codes is that their design has an additional degree of freedom: when building $\mathbf{H}$, once we fix the degrees, the only freedom that we have in the binary case concerns the choice of the positions of the $1$'s in $\mathbf{H}$; in the nonbinary case, instead, we also have to fix the values of the nonzero entries among $\{1,2,\ldots,p-1\}$. This choice plays a nontrivial role.% and we will discuss it in the next section.

\subsection{Encoding dual-diagonal LDPC codes}
The particular shape of the parity-check matrix allows to use it for encoding; given $\mathbf{H}$ as in~\eqref{eq:H} and an information vector $\mathbf{u} \in \Fp^k$, the codeword $\mathbf{c} \in \Fp^n$ associated with $\mathbf{u}$ is obtained in the following way:
\begin{enumerate}
\item $c_i = u_i$, for every $i=1,2,\ldots,k$.
\item The first parity-check equation of $C$, defined by the first equation of $\mathbf{H}$, is:
  \begin{equation*}
    \sum_{j=1}^k h_{1,j}c_j + h_{1,k+1}c_{k+1} \equiv 0 \bmod p.
  \end{equation*}
  The only unknown is $c_{k+1} \in \{0,1,\ldots,p-1\}$, that can therefore be computed easily.
\item For $i = 2,3,\ldots,n-k$, the $i$-th parity-check equation is:
  \begin{equation*}
    \sum_{j=1}^k h_{i,j}c_j + h_{i,k+i-1}c_{k+i-1} + h_{i,k+i}c_{k+i} \equiv 0 \bmod p.
  \end{equation*}
  A priori, the unknowns in the previous congruence are $c_{k+i-1}$ and $c_{k+i}$. Yet, for $i = 2$, the only unknown is $c_{k+2}$, because we computed $c_{k+1}$ in step 2). Thus, $c_{k+2}$ can be obtained too.
\item In turn, this means that we can compute $c_{k+3}$ from the third parity-check equation, and so on so forth, we recursively obtain all the $c_i$'s for $i=k+1,k+2,\ldots,n$.
\end{enumerate}
The key observation here is that, because of the sparsity of $\mathbf{L}$, most of the $h_{i,j}$'s are equal to $0$ for $i=1,2,\ldots,n-k$ and $j=1,2,\ldots,k$. More precisely, exactly $d_r$ of them are nonzero for every fixed $i$. Therefore, the number of operations required to compute $c_{k+i}$ in steps 2)-4) is constant in $n$ for every fixed $i$ and the whole encoding procedure of the LDPC code has complexity $O(d_r(n-k)) = O\left(\frac{d_rd_c}{d_r+d_c}n\right)$. We would have a higher complexity if we encoded via the generator matrix of the code, which is in general not sparse. As a consequence of the comments made in Section \ref{sec:encoding_constrA} and Section \ref{sec:leech_constellations}, the whole encoding algorithm of a Leech constellation with underlying dual-diagonal LDPC code has complexity linear in $n$.

Finally, notice that when we apply the steps 1)-4) above, the codeword $\mathbf{c}$ is built systematically, in the sense that the information vector $\mathbf{u}$ coincides with the first $k$ coordinates of $\mathbf{c}$. This guarantees that step 2) of Section \ref{sec:demapping_constrA} has no computational complexity.

\section{Simulation results}
\label{sec:simulations}
The purpose of this section is to show some numerical decoding performance of Leech constellations of dual-diagonal LDA lattices in very high dimensions. How close to capacity will we be able to get? In~\cite[Chap.~9]{Zamir2014}, Zamir considers Voronoi constellations of lattices for which the transmission scheme includes uniform dithering at the channel input and lattice decoding. Under these premises, it is shown that for high $\snr$, the gap to capacity equals the sum of the shaping loss of the coarse lattice and the coding loss of the fine lattice (where by shaping loss we mean the gap to optimal shaping and by coding loss we mean the gap to Poltyrev limit). Our coding scheme differs from the one considered by Zamir because we are not using dithering and in simulations we apply an iterative decoder instead of a lattice decoder. Nonetheless, we will see that the gap to capacity in our numerical results is consistent with the theoretical results stated in~\cite[Chap.~9]{Zamir2014}.

Now, let us start by describing the parameters of the LDA family with which we are experimenting; this corresponds to making explicit all the choices that characterize the construction of the underlying parity-check matrix $\mathbf{H}$:

\begin{itemize}
\item As already mentioned, $\mathbf{H}$ is as in $\eqref{eq:H}$. 
\item We fix $p=13$. This $p$ is ``big enough'' in a sense that will be clear later.
\item For the construction of $\mathbf{L}$, we fix $d_c = 2$ and $d_r = 1$. This is the simplest choice for a regular $\mathbf{L}$ and has the advantage of speeding the decoding procedure, because smaller degrees correspond to less edges in the associated Tanner graph and therefore to a faster iterative decoding algorithm~\cite{Ryan2009}. According to \eqref{eq:R}, the associated LDPC codes have rate $R = 1/3$. The resulting $\mathbf{H}$ is almost regular: only the first row and the last column have different degrees from the other rows and columns.
\item $\mathbf{L}$ has size $2k \times k$. We take
\begin{equation*}
  \mathbf{L} = \left(
  \begin{array}{c}
    \mathbf{\Pi_1} \\ \hline
    \mathbf{\Pi_2}
  \end{array}
  \right),
\end{equation*}
where $\mathbf{\Pi_1}$ and $\mathbf{\Pi_2}$ are two permutation matrices of size $k \times k$ chosen at random among all permutations that do not create $4$-cycles in the Tanner graph associated with $\mathbf{H}$ (hence the girth of this graph is at least $6$). %The Tanner graph is represented in Fig.~\ref{fig:tanner}.
\item The nonzero entries of $\mathbf{H}$ are optimized with the same strategy used in~\cite{diPietro2012}: for every fixed row of $\mathbf{H}$, its nonzero entries ($d_r+2 = 3$ in general or $d_r +1 =2$ for the first row) are chosen at random among all the triples (or couples for the first row) of coefficients that guarantee that the minimum Euclidean distance of the single-parity-check code defined by the corresponding parity-check equation is bigger than $\sqrt{2}$. The reader can find in~\cite[Sec.~V-A]{diPietro2012} a more detailed explanation of this technique, which has experimentally proved to yield better performance than a completely random choice of the nonzero entries of $\mathbf{H}$.
\end{itemize}

\subsection{Infinite constellations of dual-diagonal LDA lattices}
\begin{figure}
  \centering
    \includegraphics[width=0.72\columnwidth,angle=-90]{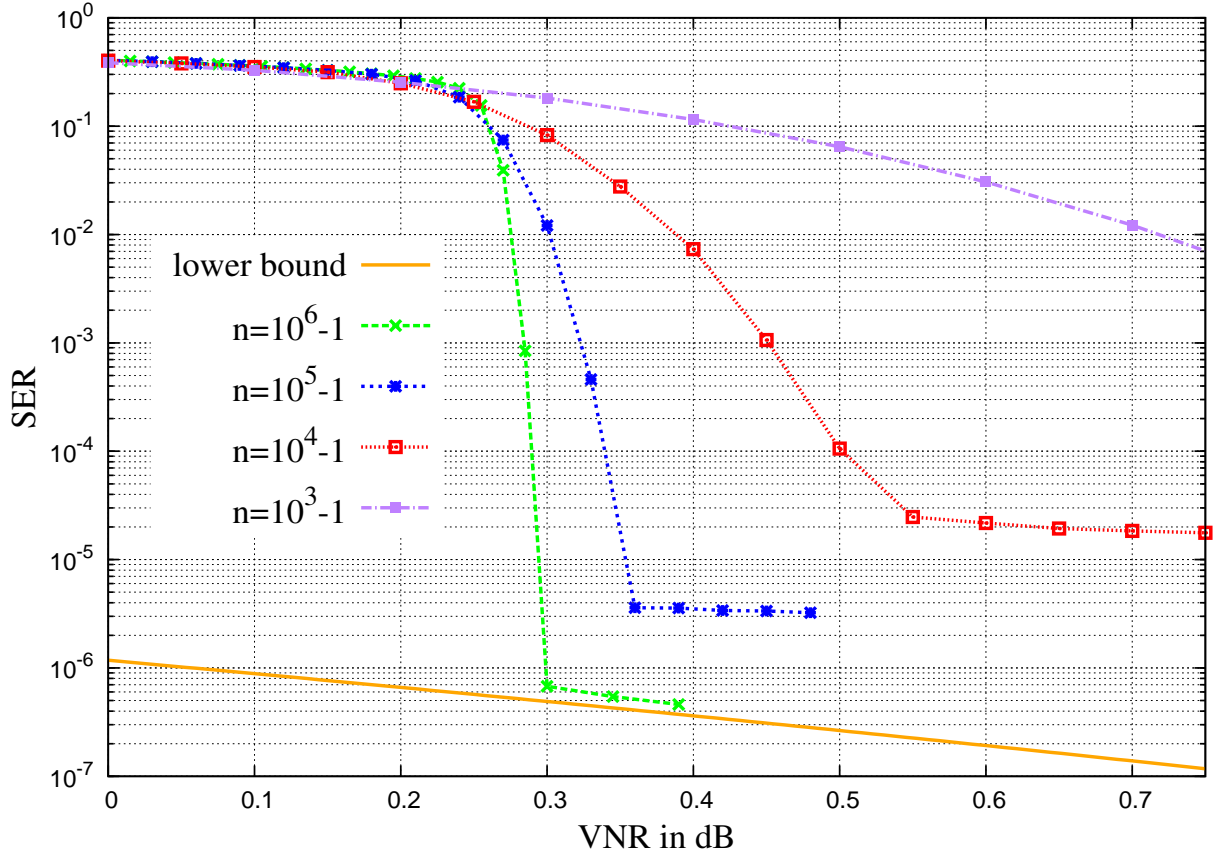}
    \caption{Performance of infinite constellations of dual-diagonal LDA lattices with parameters $p=13$, $d_c=2$, and $d_r=1$.}
    \label{fig:infinite_dd_lda}
\end{figure}

The first feature to investigate is the performance of the infinite constellations of dual-diagonal LDA lattices. Fig.~\ref{fig:infinite_dd_lda} provides some numerical evaluations of it for values of $n$ up to $10^6-1$ (notice that with our choice of the parameters $n$ has to be divisible by $3$). The decoder that we used is the same iterative belief-propagation decoder used in~\cite{diPietro2012}, whose complexity is $O(p^2n)$. Fig.~\ref{fig:infinite_dd_lda} shows the symbol-error-rate (SER) as a function of the $\vnr$. Notice that by \eqref{eq:vol_constrA}, the normalized volume of $\Lambda_f$ is $\vol(\Lambda_f)^{\frac2n} = p^{2(1-R)}$. It is clear from \eqref{eq:vnr} that fixing values of $\vnr$ bigger than $1 = 0$~dB is the same as fixing noise variances less than the Poltyrev limit $\sigma_{\max}^2$ defined in Corollary \ref{cor:poltyrev}. The waterfall region of our family of dual-diagonal LDA lattices is situated only at less than $0.3$~dB from this limit. This tells that this family is ``AWGN-good enough'' and its elements are good candidates for being the fine lattices in Leech constellations.   

Fig.~\ref{fig:infinite_dd_lda} also shows a lower bound for the SER: since $p\Zn \subseteq \Lambda_f$, the decoding performance of our LDA lattices are bounded below by those of $p\Zn$. Therefore, the decoding error probability per coordinate is bounded as follows:
  \begin{equation*}
    \prob_e(x_i) \geq \prob_e(p\Z) = 2Q\left(\frac{p}{2\sigma}\right) = 2Q\left(\sqrt{\frac{\pi e p^{2R} \vnr}{2}}\right),
  \end{equation*}
  where we obtain the last equality using \eqref{eq:vnr} and \eqref{eq:vol_constrA}.
It is interesting to notice how the SER ``touches'' the bound for $n=10^6-1$. The choice of $p=13$ is made on purpose to let the bound be less than $10^{-6}$ at $0.3$~dB from Poltyrev limit. This is the reason why we said before that $p$ is ``big enough.'' For smaller $p$, the bound would be higher in the waterfall region of Fig.~\ref{fig:infinite_dd_lda} and would not allow to fully appreciate the decoding potential of our family in the closest regions to Poltyrev limit ($\vnr = 0$~dB).

%\textcolor{blue}{The last comment to make about Fig.~\ref{fig:infinite_dd_lda} concerns the presence of the error floor which at first sight seems to decrease as $O(1/n)$. We give the following interpretation of this behavior: let us look at Fig.~\ref{fig:tanner} and focus on ......... this graph is exactly the graph of a GLD lattice as described in~\cite{Boutros2015}. In that work, we proved how small cycles are distributed in this kind of graphs and described how the error floor is caused by them, leading to symbol error probabilities of the order of $1/n$ (according to a phenomenon called \emph{spectral thinning}). Those arguments are enough to explain also the error floor of dual-diagonal LDA lattices,}
  
\subsection{Leech constellations of dual-diagonal LDA lattices}

\begin{figure}
  \centering
    \includegraphics[width=0.72\columnwidth, angle=-90]{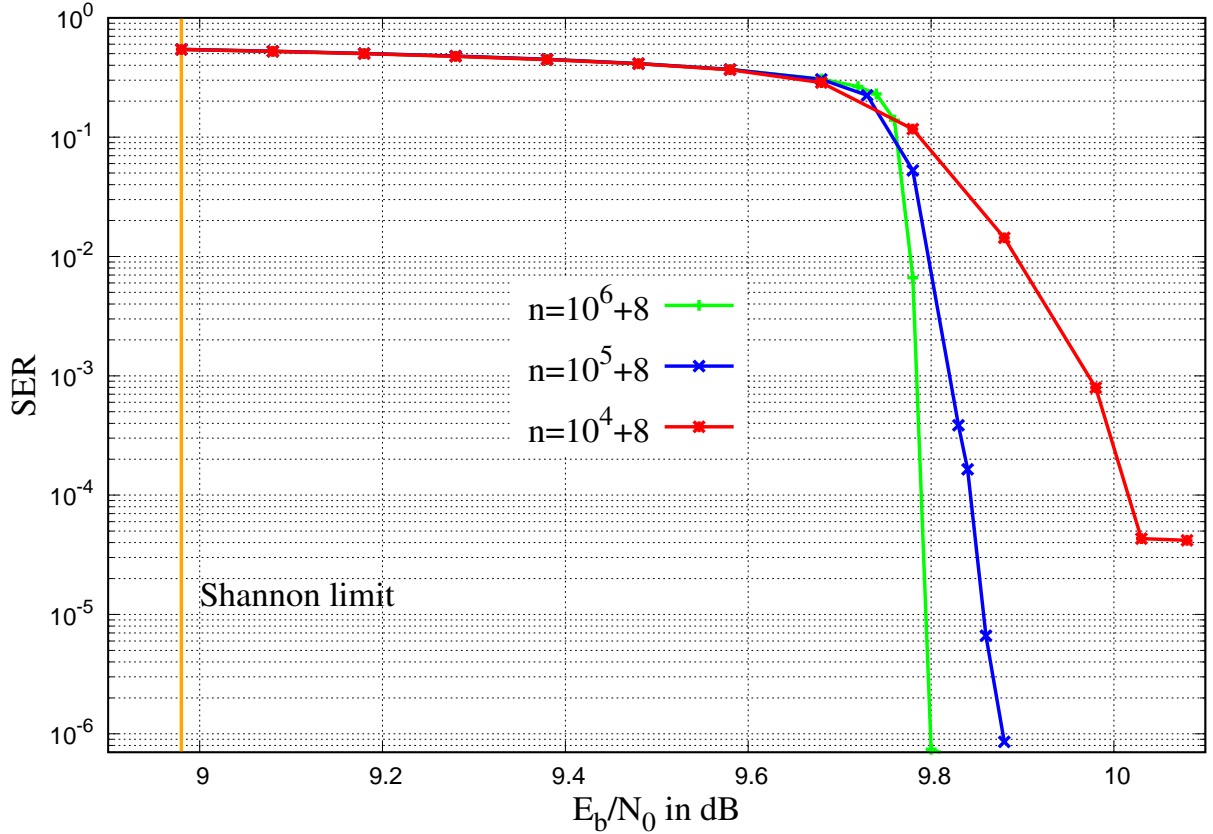}
    \caption{Performance of Leech constellations of dual-diagonal LDA lattices with information rate $R_{\mathcal{C}} \approx 2.73$ bits/dim and parameters $\alpha=1$, $p=13$, $d_c=2$, and $d_r=1$.}
    \label{fig:leech_constellation}
\end{figure}

\begin{figure}
  \centering
  \includegraphics[width=0.72\columnwidth, angle=-90]{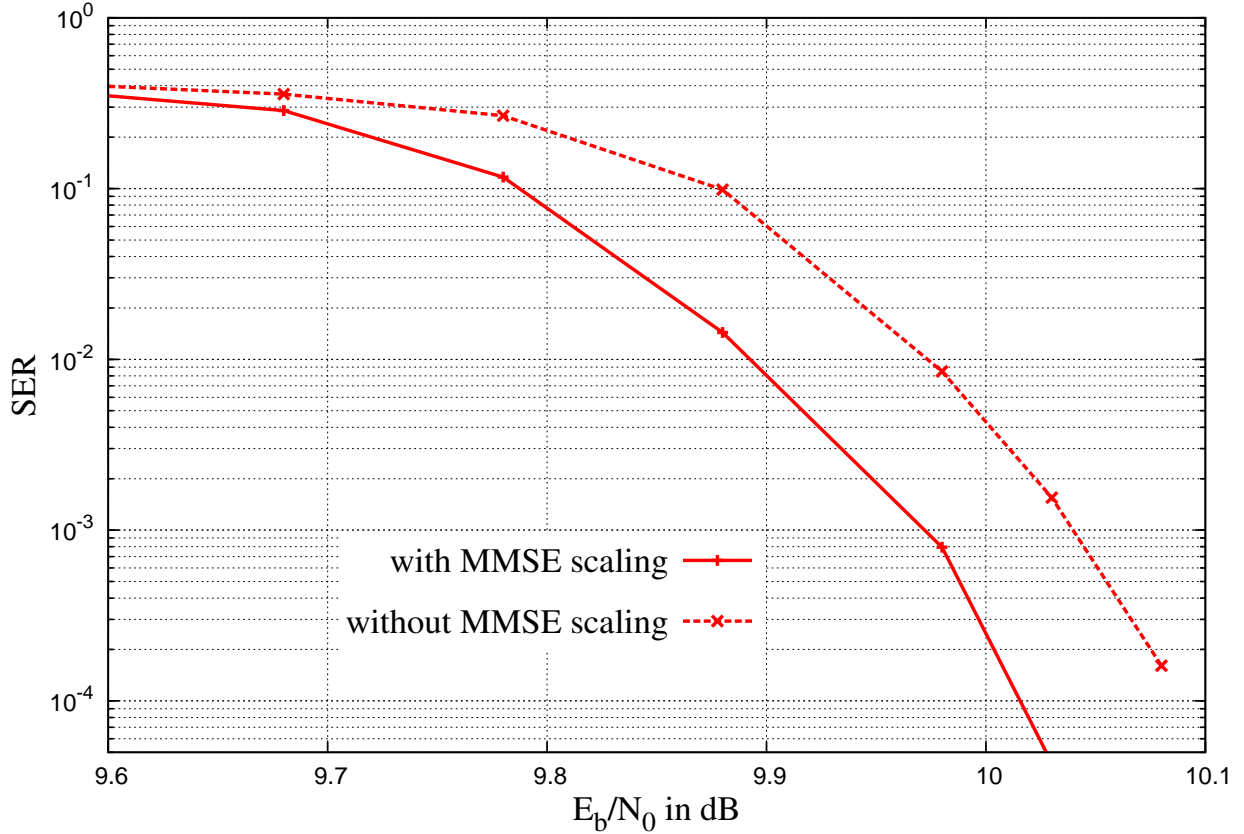}
  \caption{Performance comparison of the same Leech constellation of a dual-diagonal LDA lattice with and without MMSE scaling before decoding. The information rate is $R_{\mathcal{C}} \approx 2.73$ bits/dim and the parameters of the constellation are $n=10^4+8$, $\alpha=1$, $p=13$, $d_c=2$, and $d_r=1$.}
  \label{fig:mmse_comparison}
\end{figure}

In the previous subsection we established that our family of dual-diagonal LDA lattices has good infinite-constellation performance. Now it is time to validate the goodness of its Leech constellations too. As anticipated at the end of Section \ref{sec:leech_constellations}, with our choice of the parameters and fixing $\alpha =1$, the information rate of the constellations is $R_{\mathcal{C}} \approx 2.73$ bits per dimension. In Fig.~\ref{fig:leech_constellation}, we can see the SER numerically measured as a function of $E_b/N_0$ in dimensions up to $n=10^6+8$ (recall that in this case $n$ has to be divisible by $24$). In this scenario, Shannon capacity corresponds to $E_b/N_0 = 8.98$~dB, where, as usual, $\sigma^2 = N_0/2$ and $E_b = P/R_{\mathcal{C}}$ is the average energy per bit of our constellation.

The experimental gap to Shannon capacity shown in Fig.~\ref{fig:leech_constellation} equals $0.8$~dB. As announced at the beginning of the section, it corresponds to the sum of the gap to Poltyrev limit measured in Fig.~\ref{fig:infinite_dd_lda} (around $0.3$~dB) plus the gap to optimal shaping due to our choice of shaping the constellation with a direct sum of copies of the Leech lattice (around $0.5$~dB). %Finally, let us mention that the results of Fig.~\ref{fig:leech_constellation} are remarkable also for the following reason: let us consider for instance lattices in dimension close to $10^6$; if we assume that the infinite-constellation performance of Fig.~\ref{fig:infinite_dd_lda} for $n=10^6-1$ approximates well enough the ``typical'' behavior of double-diagonal LDA lattices in that dimension range, then the experimental waterfall region of Fig.~\ref{fig:leech_constellation} for $n=10^6+8$ is situated at a distance of around $(1.53 - \gamma_s(\Lambda) + \Delta)$~dB from the Shannon limit, where $1.53$~dB is the optimal shaping gain, $\gamma_s(\Lambda) = \gamma_s(\Lambda_{24})$ is the shaping gain \eqref{eq:shaping_gain} of our shaping lattice, and $\Delta$ is the distance from the Poltyrev limit of the waterfall region in Fig.~\ref{fig:infinite_dd_lda}. We could not hope better.

The Leech-lattice quantizer that we use to perform step 4) of encoding (cf. Section \ref{sec:encoding_constrA}) is the sphere decoder by Viterbo and Boutros~\cite{Viterbo1999}; an alternative can be the ML Leech decoder of~\cite{Vardy1993}. The fine dual-diagonal LDA lattices are decoded with the same iterative belief-propagation decoder employed for the infinite constellations of Fig.~\ref{fig:infinite_dd_lda}. However, before decoding, in this case we multiply the channel output $\mathbf{y} = \mathbf{x} + \mathbf{n}$ by the \emph{Wiener coefficient}:
\begin{equation*}
  w = \frac{\snr}{1+\snr},
\end{equation*}
with $\snr$ as in \eqref{eq:snr} and $P = E_b R_{\mathcal{C}}$. In other words, the iterative decoder input is $w\mathbf{y}$ instead of $\mathbf{y}$. The multiplication by $w$ is known as \emph{minimum mean squared error (MMSE) scaling}~\cite{Zamir2014}. The importance of MMSE scaling to achieve capacity with lattices over the AWGN channel and an optimal lattice decoder is explained in~\cite{Erez2004,Forney2004,diPietro2016}. %We do not know any result that theoretically justifies the use of MMSE scaling for improving iterative decoding, but practice shows its usefulness also in this case at small $E_b/N_0$ (notice that when $E_b/N_0$ grows, $w$ tends to $1$ and MMSE scaling affects less and less the channel output). In particular,
Its usefulness for decoding lattice codes with belief-propagation iterative decoders is explained in~\cite{Ferdinand2014bis}. Fig.~\ref{fig:mmse_comparison} shows an example of the difference in decoding the same Leech constellation with or without MMSE scaling. The constellation that we have taken into account is the same whose performance is plotted in Fig.~\ref{fig:leech_constellation} for $n = 10^4+8$. MMSE scaling before decoding allows to gain up to almost $0.1$~dB at low $E_b/N_0$. Notice that this is very close to the theoretical performance gain of $10\log_{10}(1/w)$~dB predicted by the scheme of~\cite{Erez2004}. This gain is not negligible when we work close to capacity.

\section{Conclusion}
\label{sec:conclusion}
This paper contains two main results: the description of a novel encoding and demapping scheme for Construction-A lattices and its application to design Voronoi constellations whose encoding, decoding, and demapping complexities are all linear in the lattice dimension $n$. The latter result is the combination of several factors:
  \begin{enumerate}
  \item The use of dual-diagonal LDPC codes for Construction A. Their encoding algorithm exploits the particular shape of the parity-check matrix and has linear complexity in $n$.
  \item The application of an encoding scheme for Construction-A lattices which does not require multiplication by the lattice generator matrix. This scheme is based on the new characterization of a set of coset leaders of $\Lambda_f/\Lambda$ given in Lemma \ref{lem:classes}.
  \item The use of a direct sum of copies of the Leech lattice as a shaping lattice. This choice makes demapping linear in $n$ and contributes to the linearity of the encoding complexity.
  \item The use of a low-complexity iterative decoding algorithm.
  \end{enumerate}

The effectiveness of our construction is confirmed by the numerical simulations of the previous section. To obtain that performance, we selected a very specific family of LDA lattices, optimized in many little but nontrivial senses.
%: we paid much attention to the choice of the prime number underlying Construction A, of the row and column degrees of the parity-check matrix, of the construction of its left submatrix, and of its nonzero entries. 
The result of these choices is satisfying and invites to further investigate these kinds of constructions. Notice that the encoding, decoding, and demapping procedures are independent from the choice of the shaping lattice and using the Leech lattice for shaping is  not mandatory. With the same principles applied in this paper, we can build Voronoi constellations using any Construction-A fine lattice and any integer lattice $\Gamma$ to build the coarse lattice $p\Gamma = \Lambda$. In particular, lattices in smaller dimensions can substitute the Leech lattice to improve the encoding complexity; others with better shaping gain can be used to improve the decoding performance.


\begin{thebibliography}{99}
  
\bibitem{Forney1984}
  G.~Forney, R.~Gallager, G.~Lang, F.~Longstaff, and S.~Qureshi,
  ``Efficient modulation for band-limited channels,''
  \emph{IEEE J. Sel. Areas Commun.}, vol.~2, no.~5, pp.~632-647, Sep.~1984.
  
\bibitem{Zamir2014}
  R.~Zamir,
  \emph{Lattice coding for signals and networks},
  Cambridge, United Kingdom: Cambridge University Press, 2014.  
  
\bibitem{Ma2015}
  Z.~Ma, Z.-Q.~Zhang, Z.-G.~Ding, P.-Z.~Fan, and H.-C.~Li,
  ``Key techniques for 5G wireless communications: network architecture, physical layer, and MAC layer perspectives,''
  \emph{Science China Information Sciences}, vol.~58, no.~4, pp.~1-20, Apr.~2015.
  
\bibitem{deBuda1989}
  R.~de~Buda,
  ``Some optimal codes have structure,''
  \emph{IEEE J. Sel. Areas Commun.}, vol.~7, no.~6, pp.~893-899, Aug.~1989.
  
\bibitem{Poltyrev1994}
  G.~Poltyrev,
  ``On coding without restrictions for the AWGN channel,''
  \emph{IEEE Trans. Inf. Theory}, vol.~40, no.~2, pp.~409-417, Mar.~1994.
  
\bibitem{Loeliger1997}
  H.-A.~Loeliger,
  ``Averaging bounds for lattices and linear codes,''
  \emph{IEEE Trans. Inf. Theory}, vol.~43, no.~6, pp.~1767-1773, Nov.~1997.
  
\bibitem{Urbanke1998}
  R.~Urbanke and B.~Rimoldi,
  ``Lattice codes can achieve capacity on the AWGN channel,''
  \emph{IEEE Trans. Inf. Theory}, vol.~44, no.~1, pp.~273-278, Jan.~1998.

\bibitem{Erez2004}
  U.~Erez and R.~Zamir,
  ``Achieving $\frac{1}{2}\log(1+\snr)$ on the AWGN channel with lattice encoding and decoding,''
  \emph{IEEE Trans. Inf. Theory}, vol.~50, no.~10, pp.~2293-2314, Oct.~2004.

\bibitem{Nazer2011}
  B.~Nazer and M.~Gastpar,
  ``Compute-and-forward: harnessing interference through structured codes,''
  \emph{IEEE Trans. Inf. Theory}, vol.~57, no.~10, pp.~6463-6486, Oct.~2011.

\bibitem{Ingber2013}
  A.~Ingber, R.~Zamir, and M.~Feder, 
  ``Finite-dimensional infinite constellations,''
  \emph{IEEE Trans. Inf. Theory}, vol.~59, no.~3, pp.~1630-1656, Mar.~2013.

\bibitem{Ling2014}
  C.~Ling and J.-C.~Belfiore, 
  ``Achieving AWGN channel capacity with lattice Gaussian coding,''
  \emph{IEEE Trans. Inf. Theory}, vol.~60, no.~10, pp.~5918-5929, Oct.~2014.

\bibitem{Ordentlich2016}
O.~Ordentlich and U.~Erez,
``A simple proof for the existence of ``good'' pairs of nested lattices,''
\emph{IEEE Trans. Inf. Theory}, vol.~62, no.~8, pp.~4439-4453, Aug.~2016.

\bibitem{Sadeghi2006}
  M.-R.~Sadeghi, A.~H.~Banihashemi, and D.~Panario,
  ``Low-density parity-check lattices: construction and decoding analysis,''
  \emph{IEEE Trans. Inf. Theory}, vol.~52, no.~10, pp.~4481-4495, Oct.~2006.

\bibitem{Sommer2008}
  N.~Sommer, M.~Feder, and O.~Shalvi,
  ``Low-density lattice codes,''
  \emph{IEEE Trans. Inf. Theory}, vol.~54, no.~4, pp.~1561-1585, Apr.~2008.

\bibitem{Sakzad2011}
  A.~Sakzad, M.-R.~Sadeghi, and D.~Panario,
  ``Turbo lattices: construction and error decoding performance,''
  Aug.~2011. Available: \url{http://arxiv.org/abs/1108.1873}

\bibitem{diPietro2012}
  N.~di~Pietro, J.~J.~Boutros, G.~Z\'emor, and L.~Brunel,
  ``Integer low-density lattices based on Construction A,''
  in \emph{Proc. ITW}, Lausanne, Switzerland, 2012, pp.422-426.

\bibitem{Sadeghi2013}
  M.-R.~Sadeghi and A.~Sakzad,
  ``On the performance of $1$-level LDPC lattices,''
  in \emph{Proc. IWCIT}, Tehran, Iran, 2013, pp.~1-5.

\bibitem{Yan2014}
  Y.~Yan, L.~Liu, C.~Ling, and X.~Wu,
  ``Construction of capacity-achieving lattice codes: polar lattices,''
  Nov.~2014. Available: \url{http://arxiv.org/abs/1411.0187}

\bibitem{Boutros2015}
  J.~J.~Boutros, N.~di~Pietro, and Y.-C.~Huang,
  ``Spectral thinning in GLD lattices,''
  in \emph{Proc. ITA Workshop}, La Jolla (CA), USA, 2015, pp.1-9.

\bibitem{Vatedka2017}
  S.~Vatedka and N.~Kashyap,
  ``Some ``goodness'' properties of LDA lattices,''
  \emph{Probl. Inf. Transm.}, vol.~53, no.~1, pp.~1-29, Jan.~2017.

\bibitem{diPietro2016}
  N.~di~Pietro, G.~Z\'emor, and J.~J.~Boutros,
  ``LDA lattices without dithering achieve capacity on the Gaussian channel,''
  Mar.~2016. Available: \url{http://arxiv.org/abs/1603.02863}

%\bibitem{Conway1982}
%  J.~Conway and N.~J.~A.~Sloane, 
%  ``Voronoi regions of lattices, second moments of polytopes, and quantization,''
%  \emph{IEEE Trans. Inf. Theory}, vol.~28, no.~2, pp.~211-226, Mar.~1982.

\bibitem{Conway1983}
  J.~Conway and N.~J.~A.~Sloane, 
  ``A fast encoding method for lattice codes and quantizers,''
  \emph{IEEE Trans. Inf. Theory}, vol.~29, no.~6, pp.~820-824, Nov.~1983.

\bibitem{Erez2005bis}
    U.~Erez and S.~ten~Brink,
    ``A close-to-capacity dirty paper coding scheme,''
    \emph{IEEE Trans. Inf. Theory}, vol.~51, no.~7, pp.~3417-3432, Oct.~2005.

\bibitem{Ferdinand2016}
  N.~S.~Ferdinand, B.~M.~Kurkoski, M.~Nokleby, and B.~Aazhang,
  ``Low-dimensional shaping for high-dimensional lattice codes,''
  \emph{IEEE Trans. Wireless Commun.}, vol.~15, no.~11, pp.~7405-7418, Nov.~2016.

\bibitem{Kurkoski2009}
  B.~M.~Kurkoski, J.~Dauwels, and H.-A.~Loeliger,
  ``Power-constrained communications using LDLC lattices,''
  in \emph{Proc. ISIT}, Seoul, Korea, 2009, pp.739-743.

\bibitem{Sommer2009}
  N.~Sommer, M.~Feder, and O.~Shalvi,
  ``Shaping methods for low-density lattice codes,''
  in \emph{Proc. ITW}, Taormina, Italy, 2009, pp.238-242.

\bibitem{Ferdinand2014}
  N.~S.~Ferdinand, B.~M.~Kurkoski, B.~Aazhang, and M.~Latva-aho,
  ``Shaping low-density lattice codes using Voronoi integers,''
  in \emph{Proc. ITW}, Hobart, Australia, 2014, pp.128-132.

\bibitem{Kurkoski2016}
  B.~M.~Kurkoski,
  ``Encoding and indexing of lattice codes,''
  July~2016. Available: \url{http://arxiv.org/abs/1607.03581}

\bibitem{Erez2005}
  U.~Erez, S.~Litsyn, and R.~Zamir,
  ``Lattices which are good for (almost) everything,'' 
  \emph{IEEE Trans. Inf. Theory}, vol.~42, no.~10, pp.~3401-3416, Oct.~2005.

\bibitem{Conway1999}
  J.~Conway and N.~J.~A.~Sloane, 
  \emph{Sphere packings, lattices and groups}, 3rd~ed.,
  New York (NY), USA: Springer-Verlag, 1999.

\bibitem{Ebeling2013}
  W.~Ebeling,
  \emph{Lattices and codes}, 3rd~ed.,
  Wiesbaden, Germany: Springer Spektrum, 2013. %,pp.xxx-xxx

\bibitem{Cohn2017}
  H.~Cohn, A.~Kumar, S.~D.~Miller, D.~Radchenko, and M.~Viazovska,
  ``The sphere packing problem in dimension 24,''
  \emph{Princeton Ann. Math.}, vol.~185, no.~3, pp.~1017-1033, May~2017.

\bibitem{Forney2000}
  G.~D.~Forney, Jr., M.~D.~Trott, and S.-Y.~Chung,
  ``Sphere-bound-achieving coset codes and multilevel coset codes,''
  \emph{IEEE Trans. Inf. Theory}, vol.~46, no.~3, pp.~820-850, May~2000.

\bibitem{diPietro2013}
  N.~di~Pietro, J.~J.~Boutros, G.~Z\'emor
  ``New results on Construction A lattices based on very sparse parity-check matrices,''
  in \emph{Proc. ISIT}, 2013, Istanbul, Turkey, pp.1675-1679.

\bibitem{diPietro2015}
  N.~di~Pietro, N.~Basha, and J.~J.~Boutros,
  ``Non-binary GLD codes and their lattices,''
  in \emph{Proc. ITW}, Jerusalem, Israel, 2015, pp.1-5.

\bibitem{Proakis2008}
  J.~G.~Proakis and M.~Salehi,
  \emph{Digital communications}, 5th~ed.,
  New York (NY), USA: McGraw-Hill, 1996.

\bibitem{Forney1989}
  G.~D.~Forney, Jr.,
  ``Multidimensional constellations. II. Voronoi constellations,''
  \emph{IEEE J. Sel. Areas Commun.}, vol.~7, no.~6, pp.~941-958, Aug.~1989.

\bibitem{Forney1998}
  G.~D.~Forney, Jr. and G.~Ungerboeck,
  ``Modulation and coding for linear Gaussian channels,''
  \emph{IEEE Trans. Inf. Theory}, vol.~44, no.~6, pp.~2384-2415, Oct.~1998.
  
\bibitem{Cohen1996}
  H.~Cohen,
  \emph{A course in computational algebraic number theory}, 3rd~ed.,
  Berlin, Heidelberg, Germany: Springer-Verlag, 1996.
  
\bibitem{Forney1992}
  G.~D.~Forney, Jr.,
  ``Trellis shaping,''
  \emph{IEEE Trans. Inf. Theory}, vol.~38, no.~2, pp.~281-300, Mar.~1992.
  
  %To equalize the last page columns:
  %\newpage

\bibitem{Gallager1963}
  R.~G.~Gallager, 
  \emph{Low-density parity-check codes},
  Cambridge (MA), USA: MIT Press, 1963.

\bibitem{Ryan2009}
  W.~E.~Ryan and S.~Lin,
  \emph{Channel codes: classical and modern}, %5th~ed.,
  New York (NY), USA: Cambridge University Press, 2009.
  
%\bibitem{Guo2010}
%  Q.~Guo-lei and D.~Zi-jian,
% ``Design of structured LDPC codes with quasi-cyclic and rotation architecture,''
%  in \emph{Proc. ICACTE}, Chengdu, China, 2010, pp.V1-655-V1-657.

%\bibitem{He2006}
%  Z.~He, P.~Fortier, and S.~Roy,
%  ``A class of irregular LDPC codes with low error floor and low encoding complexity,''
%  \emph{IEEE Commun. Lett.}, vol.~10, no.~5, pp.~372-374, May~2006.

%\bibitem{He2014}
%  Z.~He, P.~Fortier, S.~Roy, and H.~Xu,
%  ``An encoder/decoder with throughput over Gigabits/sec for rate-compatible LDPC codes with wide code rates,''
%  in \emph{Proc. NEWCAS}, Trois-Rivi\`eres (QC), Canada, 2014, pp.181-184.

%\bibitem{Li2014}
%  W.~Li, B.~Chen, J.~Lei, and E.~Li,
%  ``Low density parity check codes with quasi-cyclic structure and zigzag pattern,''
%  in \emph{Proc. ICSP}, HangZhou, China, 2014, pp.1-5.

%\bibitem{Schmalen2012}
%  L.~Schmalen, S.~ten~Brink, G.~Lechner, and A.~Leven,
%  ``On threshold prediction of low-density parity-check codes with structure,''
%  in \emph{Proc. CISS}, Princeton (NJ), USA, 2012, pp.1-5.

\bibitem{Viterbo1999}
  E.~Viterbo and J.~Boutros,
  ``A universal lattice code decoder for fading channels,''
  \emph{IEEE Trans. Inf. Theory}, vol.~45, no.~5, pp.~1639-1642, July~1999.

\bibitem{Vardy1993}
  A.~Vardy and Y.~Be'ery,
  ``Maximum likelihood decoding of the Leech lattice,''
  \emph{IEEE Trans. Inf. Theory}, vol.~39, no.~4, pp.~1435-1444, July~1993.

\bibitem{Forney2004}
  G.~D.~Forney, Jr.,
  ``On the role of MMSE estimation in approaching the information-theoretic limits of linear Gaussian channels: Shannon meets Wiener,''
  in \emph{Proc. Commun., Control, and Computing, 2003 41st Annu. Allerton Conf. on},
  Monticello (IL), USA, 2003, pp.~1-14.

\bibitem{Ferdinand2014bis}
  N.~S.~Ferdinand, M.~Nokleby, B.~M.~Kurkoski, and B.~Aazhang,
  ``MMSE scaling enhances performance in practical lattice codes,''
  in \emph{Proc. ACSSC}, Pacific Grove (CA), USA, 2014, pp.1021-1025.

\end{thebibliography}
\end{document}